\documentclass[aip,jmp
]{revtex4}
\usepackage{graphicx}% Include figure files
\usepackage{dcolumn}% Align table columns on decimal point
\usepackage{bm}% bold math
\usepackage{amsmath}
\usepackage{amssymb}
\usepackage{amsthm}

\newcommand{\tr}{\mathop{\mathrm{tr}}\nolimits}
\renewcommand{\Re}{\mathop{\mathrm{Re}}\nolimits}
\renewcommand{\Im}{\mathop{\mathrm{Im}}\nolimits}
\newcommand{\RM}{\mathbb{R}}

\newcommand{\NM}{\mathbb{N}}

\newtheorem{theorem}{Theorem}
\newtheorem{lemma}{Lemma}
\newtheorem{corollary}{Corollary}
\newtheorem{proposition}{Proposition}

\theoremstyle{definition}
\newtheorem{remark}{Remark}
\newtheorem{example}{Example}

\begin{document}

%\preprint{}

\title{Quantum Zeno effect and dynamics} %Title of paper



\author{Paolo Facchi}
%\thanks{}
\affiliation{Dipartimento di Matematica, Universit\`a di Bari, I-70125 Bari, Italy}
\affiliation{Istituto Nazionale di Fisica Nucleare, Sezione di Bari, I-70126 Bari, Italy}

\author{Marilena Ligab\`o}
%\thanks{}
\affiliation{Dipartimento di Matematica, Universit\`a di Bari, I-70125 Bari, Italy}


\date{\today}

\begin{abstract}
If frequent measurements ascertain whether a quantum system is still in a given subspace, it remains in that subspace and a quantum Zeno effect takes place. The limiting time evolution within the projected subspace is called quantum Zeno dynamics. This phenomenon is related to the limit of a product formula obtained by intertwining the time evolution group with an orthogonal projection.

By introducing a novel product formula we will give a characterization of the  quantum Zeno effect for finite-rank projections, in terms of a spectral decay property of the  Hamiltonian in the range of the projections.
Moreover, we will also characterize its limiting quantum Zeno dynamics and exhibit its (not necessarily lower-bounded) generator as a generalized mean value Hamiltonian.
\end{abstract}


\maketitle %\maketitle must follow title, authors, abstract and \pacs


\section{\label{sec:intro}Introduction}

Frequent measurements can slow down the evolution of a quantum system and eventually
hinder any transition to states different from the initial one. This
phenomenon, first considered by Beskow and Nilsson~\cite{beskow} in their
study of the decay of unstable systems,
was  named quantum Zeno
effect by Misra and Sudarshan~\cite{misra},
who suggested a parallelism with the paradox of the arrow by the
philosopher Zeno of Elea.

Since then,  the quantum Zeno effect has received constant attention by physicists and mathematicians, who explored different facets of the phenomenon. The whole field is very active. For an up-to-date review of the main mathematical and physical aspects, see \cite{ZenoMP} and references therein.

The  quantum Zeno effect has been observed  experimentally in a variety of systems, on
experiments involving photons, nuclear spins, ions,  optical pumping, photons in a cavity, ultracold atoms and Bose-Einstein condensates. In all the above-mentioned implementations, the quantum system is forced to remain in its
initial state through a measurement associated to a one-dimensional projection. No experiment has been performed so far in order to check the multi-dimensional quantum Zeno effect and the quantum Zeno dynamics, that is the effective limiting dynamics inside the measured subspace. However, these ideas might lead to remarkable applications, e.g.\ in quantum computation and in the control of decoherence.

From the mathematical point of view the quantum Zeno dynamics is related to the
limit of a product formula obtained by intertwining the dynamical time evolution group  with the orthogonal projection associated with the measurements performed on the system. It can be viewed as a generalization of Trotter-Kato product formulae \cite{Trotter1,Trotter2,Kato1,Chernoff}  to more singular objects in which one semigroup is replaced by a  projection.

Since the seminal paper by Misra and Sudarshan \cite{misra}, the
main object of interest has been the limit of the
quantum Zeno dynamics product formula.
Its structure has been thoroughly investigated and has been well characterized under quite general
hypotheses. In particular, by assuming that  the Hamiltonian is lower bounded and the limit is strongly continuous, one obtains  a unitary group within the projected subspace \cite{misra,Exnerbook}.

On the other hand, the much more difficult question
of the existence of this limit, for infinite dimensional projections and unbounded Hamiltonian,
is still open.
Since this product formula and its properties are of great
importance in the  study of quantum dynamical semigroups and
have remarkable consequences both in mathematical
physics and operator theory, there have been many investigations
from different perspectives and motivations. See, for example, \cite{Friedman72,Friedman1,gustafsonmisra,Schmidt02,Schmidt03,Matolcsi03,EINZ}.

 In 2005 Exner and Ichinose~\cite{exner} proved the existence of a quantum Zeno dynamics when the Hamiltonian is positive and the domain of its square root has a dense intersection with the range $\mathcal{H}_P$ of the projection. However, this result was proved in the $L^2_{\textrm{loc}}(\RM,\mathcal{H}_P)$ topology, instead of the more natural strong operator topology.
As a corollary of the main result, they
solved the problem in the norm operator topology when the projections are finite dimensional.

The first main result that we present in this paper is a complete characterization
of the multi-dimensional quantum Zeno effect, for Hamiltonians that are not
necessarily lower bounded, through the introduction of a novel
product formula directly related to the quantum Zeno effect.
 We show that the existence of the limit is related to
a fall-off property of the spectral measure of the Hamiltonian in
the range of the projection.

Then, we also exhibit a characterization of the quantum Zeno dynamics, in terms of the above-mentioned energy fall-off property and of the existence of a mean value Hamiltonian in a generalized sense.

This paper is organized as follows. In section \ref{sec:results} we discuss the relation
between the quantum Zeno effect and its limiting dynamics, in particular
we recall the product formula related to the quantum Zeno dynamics.
Then, we introduce a new product formula which is
directly related to the quantum Zeno effect and present our first theorem on the
characterization of the existence of its limit.
Finally, in the second theorem, we will give  a characterization of the related quantum Zeno dynamics.
Moreover,  we consider  an
example that explains  the differences between the
conditions that imply the quantum Zeno dynamics and the quantum Zeno effect.
The proofs of the theorems are postponed to section \ref{sec:iff condition}.

\section{\label{sec:results}Quantum Zeno effect vs quantum Zeno dynamics. Results}

Consider a quantum system $Q$, whose states are described by density operators, that are positive operators with unit trace, in a complex separable Hilbert space
$\mathcal{H}$.
The time evolution of the system is governed by a unitary group
$U(t)=\exp(-itH)$, where $H$ is a time-independent self-adjoint Hamiltonian.
Consider also an orthogonal projection $P$, that describes the
measurement process that is performed on $Q$.
This kind of measurement ascertains whether the system is in the
subspace $\mathcal{H}_{P}:=P\mathcal{H}$.
Assume that the initial
density operator $\rho_{0}$
of the system has support in $\mathcal{H}_{P}$, namely
\begin{equation*}
\rho_{0}=P\rho_{0}P, \quad \tr(\rho_{0}P)=1.
\end{equation*}
The state of the system at time $\tau$ is
\begin{equation*}
\rho(\tau)=U(\tau)\rho_{0}U(\tau)^{*}
\end{equation*}
and after a measurement, if the outcome is positive, it becomes
\[
\frac{P\rho(\tau)P}{p(\tau)}=\frac{V(\tau)\rho_{0}V(\tau)^{*}}{p(\tau)}
\]
where $V(\tau)=PU(\tau)P$, and $p(\tau)=
\tr(V(\tau)\rho_{0}V(\tau)^{*})$. Observe that, since $P$ is not
assumed to commute with the Hamiltonian, when $[P,H]\neq 0$, the
unitary evolution drives the system outside $\mathcal{H}_{P}$, and
$p(\tau)$ is in general smaller than unity.

If we perform a series of $P$-observations on $Q$ at time
$\tau_{j}=jt/N$, $j \in \{1, \ldots, N\}$, its state after $N$
positive measurements is, up to a normalization,
\[
\rho_{N}(t)=V_{N}(t)\rho_{0}V_{N}(t)^{*}
\]
where $V_{N}(t)=(PU(t/N)P)^N$ and the survival probability in
$\mathcal{H}_{P}$ reads
\begin{equation}
\label{eq:pN(t)}
p_{N}(t)= \tr (V_{N}(t)\rho_{0}V_{N}(t)^{*}).
\end{equation}
Our interest is focused on the following question: under what
conditions
\begin{equation}
p_{N}(t) \to 1, \qquad \text{for} \quad N \to +\infty?
\label{eq:QZEdef}
\end{equation}
Misra and Sudarshan \cite{misra}  baptized this problem \emph{quantum Zeno effect} (QZE): repeated $P$-observations in succession inhibit transitions
outside the observed subspace $\mathcal{H}_{P}$. That is, rephrasing the Greek philosopher Zeno, the observed quantum arrow does not move.

Since the seminal paper \cite{misra}, the
main object of interest has been the limit of the
following product formula
\begin{equation}\label{ampl. form}
V_{N}(t)=(PU(t/N)P)^N ,
\end{equation}
and in particular whether $U_Z(t)= \lim_N V_N(t)$  exists and is given by a unitary group in
$\mathcal{H}_{P}$. The existence of a unitary limit is tantamount to the presence of a \emph{quantum Zeno dynamics} (QZD).
If this is the case, one immediately gets
\begin{equation*}
\lim_{N\to\infty} p_N(t)= \tr (U_Z(t)\rho_{0}U_Z(t)^{*})= \tr (P\rho_{0})=1,
\end{equation*}
by the cyclic property of the trace. Namely, QZD implies QZE.

The following theorem, due to Exner and Ichinose \cite{exner}, about the existence of the limit of the QZD product formula (\ref{ampl. form}) when $P$ is a finite rank projection and $H$ is positive, provides a sufficient condition for the quantum Zeno effect. For a simple proof of this result see \cite{EINZ}.

\begin{theorem}[Exner-Ichinose \cite{exner}]
\label{th:EI}
Let $\mathcal{H}$ be a complex Hilbert space and $H$ a
positive self-adjoint operator with dense domain $D(H)\subset \mathcal{H}$.
Let $P$ be an orthogonal finite-rank projection onto $\mathcal{H}_{P}=P\mathcal{H}$. If $\mathcal{H}_{P} \subset D(H^{1/2})$, where  $D(H^{1/2})$  is the domain of the square root of $H$,  then
\[
\lim_{N \to +\infty} V_{N}(t)= P\exp\left(-it(H^{1/2}P)^{*}(H^{1/2}P)\right) ,
\]
uniformly for $t$ in finite intervals of $\RM$.
\end{theorem}

The hypothesis $\mathcal{H}_{P} \subset D(H^{1/2})$ on the pair Hamiltonian--projection can be
regarded as a condition on the spectral measure of $H$ over the
range of $P$ in the following way: for every $\psi\in\mathcal{H}$ one gets
\begin{eqnarray}\label{mean value - spectral measure}
\left\langle H\right\rangle_{P\psi} := (H^{1/2}P \psi,H^{1/2}P \psi)
= \int_{[0, +\infty)} 
\lambda \; d(P \psi, P_{\lambda}^H P \psi) < +\infty ,
\end{eqnarray}
where $\{P_{\Omega}^H\}$ is the projection-valued measure associated
to $H$.

Therefore, the above result can be summarized as follows: whenever a positive Hamiltonian
has a finite mean value  $\langle H \rangle$ on vector states in the range of $P$, frequently
$P$-observations force the state of the system to remain in the
subspace $\mathcal{H}_{P}$ and the limiting dynamics in this space is given by
the unitary group $U_{Z}(t)=P \exp(-it(H^{1/2}P)^*(H^{1/2}P))$. As a consequence,
finite-energy states 
exhibit a quantum Zeno effect. One can ask if the sufficient condition (\ref{mean value - spectral measure}) is also necessary for the
quantum Zeno effect. We will show that the answer to this question is negative. Indeed,
the QZE implies a  condition weaker than (\ref{mean value - spectral
measure}) on the spectral measure of $H$.

The first result of this paper is a characterization of the multi-dimensional quantum Zeno
effect. In order to achieve our goal we look at the problem from a different perspective. Instead of considering the product formula (\ref{ampl. form}), let us move back our attention to Eqs.~(\ref{eq:pN(t)}) and (\ref{eq:QZEdef}). By invoking the cyclic property of the trace, we will study the limit of the following product formula
\begin{equation}\label{NZPF}
Z_N(t)=V_{N}(t)^{*}V_{N}(t).
\end{equation}
One gets that the quantum Zeno effect (\ref{eq:QZEdef}) takes place if and only if
\begin{equation}
\label{eq:limit NZPF}
Z_N(t) \to P, \qquad \text{for} \quad N\to +\infty .
\end{equation}
We will call (\ref{NZPF}) \emph{QZE product formula}, as opposed to the QZD product formula (\ref{ampl. form}).

The next theorem is on the equivalence between the quantum Zeno effect and a certain fall-off condition on the spectral measure associated to the Hamiltonian,  that is weaker than (\ref{mean value
- spectral measure}).
Let us denote, as usual, with
$o(s)$ an operator-valued function defined in a neighbourhood of $0$ and such that
$\|o(s)\|/s =0$, for $s\to 0$. Let us also use the notation $A^c = \RM\setminus A$ for any subset $A\subset\RM$.

\begin{theorem}\label{iff condition QZE multidimensional}
Consider  a self-adjoint operator  $H$ and an orthogonal
finite-rank projection $P$ in a complex Hilbert space
$\mathcal{H}$. Let  $\{P_{\Omega}^H\}$ be the projection-valued
spectral measure  of $H$ and $\{U(t)=e^{-itH}\}_{t \in \RM}$ the
one-parameter unitary group generated by $H$. Consider the product
formula $Z_N(t)=V_{N}(t)^{*}V_{N}(t)$, where
$V_{N}(t)=(PU(t/N)P)^{N}$ with $t \in \RM$ and $N \in \NM^*$. The
following statements are equivalent:
\begin{enumerate}
    \item \label{fdtc}
    \begin{equation*}
      P\, P_{(-\Lambda,\Lambda)^c}^H \, P= o\left( \frac{1}{\Lambda} \right), \qquad \mathrm{for} \quad \Lambda \to + \infty;
       \end{equation*}
    \item
      \begin{equation*}
        \left. \frac{d}{ds} Z_1(s) \right|_{s=0}=0;
      \end{equation*}
    \item
    \begin{equation*}
      \lim_{N \to +\infty} Z_{N}(t)=P,
    \end{equation*}
    uniformly for $t$ in finite intervals of $\RM$.
\end{enumerate}
\end{theorem}

Two comments are now in order. First, by taking the matrix element of Condition \ref{fdtc}. one obtains that for any $\psi\in\mathcal{H}$
\[
(\psi, P
P_{(-\Lambda,\Lambda)^c}^H P\psi) 
= \int_{(-\Lambda,\Lambda)^c}  d (P\psi,P^H_\lambda P\psi) = o\left( \frac{1}{\Lambda} \right),
\]
that is
\begin{equation}
\label{eq:falloff}
\Lambda \int_{(-\Lambda,\Lambda)^c}  d (P\psi,P^H_\lambda P\psi)
\to 0, \qquad \text{for}\quad \Lambda\to +\infty.
\end{equation}
Let us compare (\ref{eq:falloff}) with condition (\ref{mean value - spectral measure}). Under the hypotheses of Theorem \ref{th:EI}, namely, $H\geq0$ and $\mathcal{H}_P\subset D(H^{1/2})$, one gets
\begin{eqnarray*}
\Lambda \int_{(-\Lambda,\Lambda)^c}  d (P\psi,P^H_\lambda P\psi) &=&
\Lambda \int_{[\Lambda,+\infty)}  d (P\psi,P^H_\lambda P\psi)
\\
&\leq&  \int_{[\Lambda,+\infty)} \lambda\; d (P\psi,P^H_\lambda P\psi) \to 0,
\end{eqnarray*}
when $\Lambda\to+\infty$. Therefore, condition (\ref{eq:falloff}) is implied by (\ref{mean value - spectral measure}), but it is weaker than the latter.

Second, when the measurement projection is
one-dimensional, we can write $P=\psi
(\psi,\cdot)=|\psi\rangle\langle\psi|$, for some
$\psi\in\mathcal{H}$ and $\|\psi\|=1$. Physically, this projection checks whether the system is in the pure state $\psi$. In this case we get
\begin{equation}
\label{eq:VsAs}
V(s)=P U(s) P=\mathcal{A}(s) P,
\end{equation}
where
\[
 s \in \RM \mapsto \mathcal{A}(s) = (\psi, U(s) \psi)= (\psi, e^{-i H s}\psi)
\]
is the \emph{survival probability amplitude} in the state $\psi$.
Its associated probability is
\[
 s \in \RM \mapsto p(s)=|\mathcal{A}(s)|^2,
\]
and represents the probability of finding  at time $s$ in state $\psi$ a system that started in $\psi$ at time $0$.

Note that the survival amplitude can be rewritten as
\[
\mathcal{A}(s) =
\int_{\RM} e^{-is\lambda}\; d\mu_{\psi}^H(\lambda),
\]
where $\mu_{\psi}^H(\Omega)=(\psi, P^H_{\Omega}\psi)$, for every Borel set $\Omega\subset\RM$, is the spectral measure of $H$ at $\psi$. Therefore, $\mathcal{A}(s)$ is nothing but the Fourier transform of the spectral measure $\mu_{\psi}^H$, i.e.\ a characteristic function, in probabilistic jargon.

Since $V_{N}(t)=[V(t/N)]^N$,
the QZE product formula (\ref{NZPF}) reads
\begin{equation*}
Z_N(t)= V_{N}(t)^{*}V_{N}(t)= [p(t/N)]^N P.
\end{equation*}
Thus for one-dimensional projections the occurrence of the  quantum Zeno effect (\ref{eq:limit NZPF}) is equivalent to the limit of the survival probability
\begin{equation}
\label{eq:QZE1dim}
[p(t/N)]^N \to 1, \qquad \text{for} \quad N \to +\infty.
\end{equation}
Physically, (\ref{eq:QZE1dim}) asserts that the system stays frozen in the initial state.

The following proposition, stated for a generic Borel probability measure on $\RM$, gives a characterization of the limit (\ref{eq:QZE1dim}) in terms of the fall-off property of the spectral measure $\mu_{\psi}^H$ for large energy values. The proof makes use of the equivalent condition of vanishing derivative of the survival probability at $s=0$.
Interestingly enough, the first step in the proof of our main Theorem \ref{iff condition QZE multidimensional} is Proposition \ref{iff condition QZE} (which is a special case of the first!)

\begin{proposition}\label{iff condition QZE}
Let $\mu$ be a Borel measure on   $\RM$, with $\mu(\RM)=1$. Define for every $s \in \RM$
\begin{equation*}
\mathcal{A}(s)=\int_{\RM} e^{-is\lambda}\; d\mu(\lambda)
\end{equation*}
and
\begin{equation*}
p(s)=|\mathcal{A}(s)|^2.
\end{equation*}
Then the following assertions are equivalent:
\begin{enumerate}
    \item \label{fdtc 1D}
    \begin{equation*}
          \mu((-\Lambda,\Lambda)^c) =o\left( \frac{1}{\Lambda} \right), \quad \mathrm{for} \quad \Lambda \to + \infty;
    \end{equation*}
    \item
      \begin{equation*}
        p'(0)=0;
      \end{equation*}
    \item \label{prod. formula 1D}
    \begin{equation*}
      \lim_{N \to +\infty} [p(t/N)]^N=1,
    \end{equation*}
    uniformly for $t$ in finite intervals of $\RM$.
\end{enumerate}
\end{proposition}

\begin{remark}
Let $\mu$ be a Borel measure on $\RM$ with $\mu(\RM)=1$. 
Suppose that $\mu$ satisfies one of the conditions of Proposition \ref{iff condition QZE}. 
Observe that for all $s \in \RM$
$$
p'(s)=\mathcal{A}'(s)\overline{\mathcal{A}(s)}+\mathcal{A}(s)\overline{\mathcal{A}'(s)},
$$
therefore
$$
p'(0)=2\Re \mathcal{A}'(0)=\lim_{s \to 0} 2 \left(\frac{\Re \mathcal{A}(s)-1}{s}\right)=\lim_{s \to 0} \frac{2}{s} \int_{\RM} (\cos ( \lambda s) -1)\; d\mu(\lambda).
$$
Then the real part of $\mathcal{A}'(0)$, that can be rewritten as
\[
\Re \mathcal{A}'(0)=- \lim_{s \to 0} \frac{2}{s} \int_{\RM} \sin^2\left(\frac{\lambda s}{2}\right)\; d\mu(\lambda),
\]
must equal $0$, while there are no constraints on the imaginary part of $\mathcal{A}'(0)$ given by
\[
\Im \mathcal{A}'(0)=- \lim_{s \to 0} \int_{\RM} \frac{\sin(\lambda s)}{s}\; d\mu(\lambda).
\]
We will show that it can also diverge. \qed
\end{remark}

\begin{example}
\label{ex:part}
Let $a>1$. We consider as $\mu$ the following probability measure
\begin{equation}
\label{eq:muexamp}
\mu(E)= a \log a \int_{E \cap [a, +\infty)} \frac{1+\log \lambda}{\lambda^2 \log^2 \lambda} \; d\lambda
\end{equation}
for every Borel set $E\subset\RM$.

\begin{figure}
\centering
\resizebox{0.75\textwidth}{!}{\includegraphics{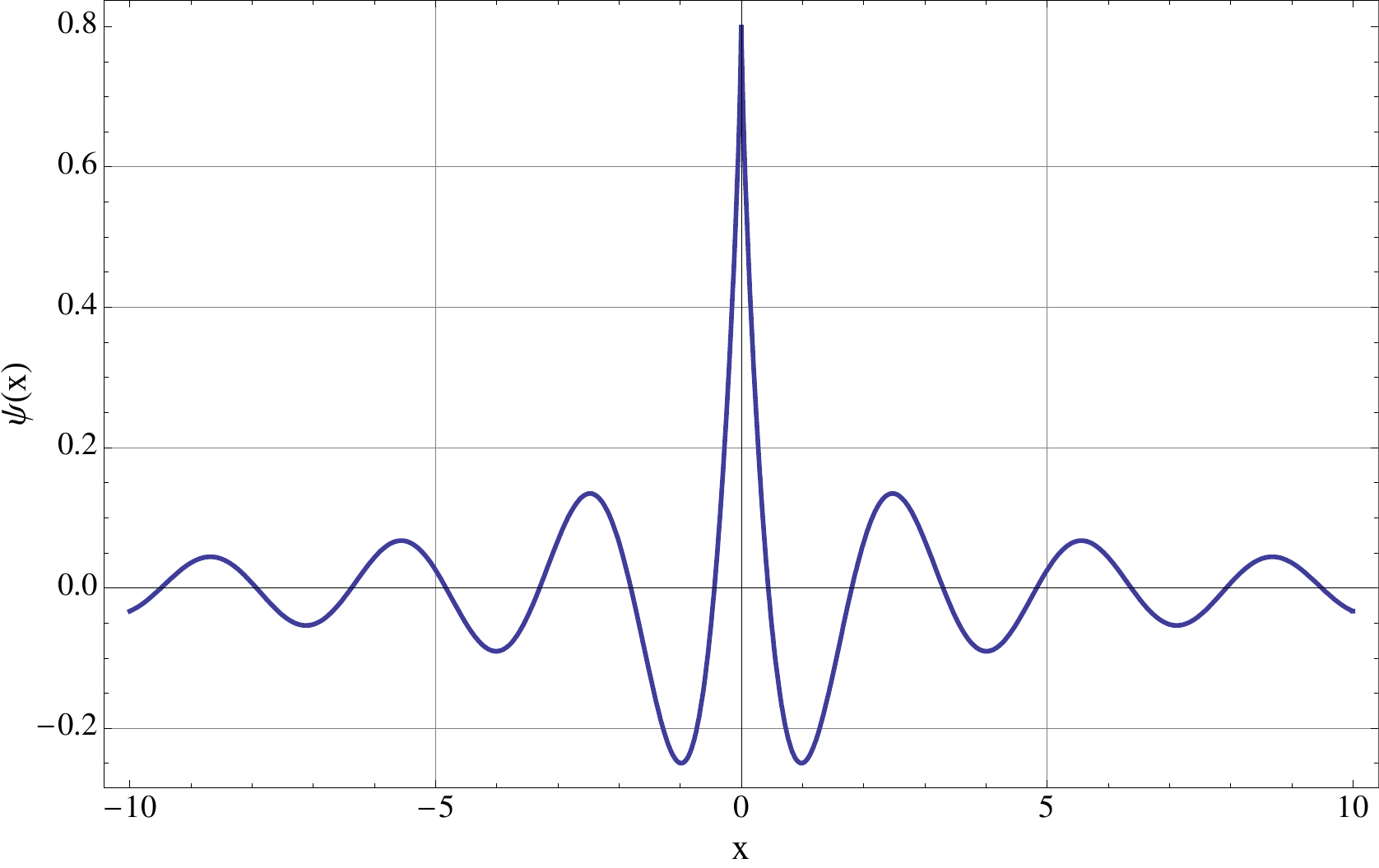}
}
%\vspace{5cm}       % Give the correct figure height in cm
\caption{Wave packet $\psi(x)$ with the fall-off property~\ref{fdtc 1D} of Proposition ~\ref{iff condition QZE}, but with infinite mean energy.}
\label{fig:1}       % Give a unique label
\end{figure}

Physically, one can implement the above example with a free particle in $n$ dimension subjected to a suitable one-dimensional projective measurement. Indeed, consider the free Hamiltonian  of a particle (with mass $m=1/2$)  $H=-\Delta$ with domain $H^2(\mathbb{R}^n)\subset L^2(\mathbb{R}^n)$.  Consider a projection $P=\psi(\psi,\cdot)=|\psi\rangle \langle\psi|$ over the (radially symmetric) wave function  $\psi\in L^2(\mathbb{R}^n)$, whose Fourier transform reads for $p\in \mathbb{R}^n$, $|p|>a^{1/2}>1$
\[
\hat{\psi}(p)=\sqrt{\frac{2 a\log a }{|S^{n-1}|}}\frac{\sqrt{1+\log |p|^2}}{|p|^{n/2+1} \log |p|^2},
\]
and $\hat{\psi}(p)=0$ otherwise,
where $|S^{n-1}|$ is the area of the unit sphere. The wave packet $\psi(x)$ for $n=1$ is plotted in Fig.\ \ref{fig:1}.
By using Fourier transform, it is not difficult to show that the spectral measure of the free Hamiltonian at state $\psi$ yields exactly the measure $\mu_\psi^H=\mu$ in (\ref{eq:muexamp}), which satisfies the hypothesis of Theorem \ref{iff
condition QZE}, because $\mu(\RM)=1$ and
\[
\lim_{\Lambda \to +\infty} \Lambda\, \mu((-\Lambda, \Lambda)^c)= \lim_{\Lambda \to +\infty} \Lambda a \log a \int_{\Lambda}^{+\infty} \frac{1+\log \lambda}{\lambda^2 \log^2 \lambda} \; d\lambda= \lim_{\Lambda \to +\infty} \frac{a\log a}{\log \Lambda}=0.
\]
However, observe that
\begin{eqnarray*}
\int_{\RM} \lambda \; d\mu(\lambda) & = & a\log a\int_{a}^{+\infty} \frac{1+\log \lambda}{\lambda \log^2 \lambda} \; d\lambda =
                                     a\log a\int_{\log a}^{+\infty} \frac{1+z}{z^2} \; dz 
                                                                       =  + \infty.
\end{eqnarray*}
Therefore, despite the fact that $\psi$ does not belong to $D(H^{1/2})=H^1(\mathbb{R}^n)$ and thus has infinite energy, by a Zeno limit one can freeze its dynamics in the initial state.

Now we show that in this case Im$\,\mathcal{A}'$ diverges. In fact, Im$\,\mathcal{A}$ has a cusp at the origin. Observe that
\begin{equation}\label{ld}
\lim_{s \to 0^+}\int_{\RM}  \frac{\sin(\lambda s)}{s}\; d\mu(\lambda)  = \lim_{s \to 0^+}\int_{|\lambda|\leq 1/s}  \frac{\sin(\lambda s)}{s}\; d\mu(\lambda)+\lim_{s \to 0^+} \int_{|\lambda| > 1/s}  \frac{\sin(\lambda s)}{s}\; d\mu(\lambda).
\end{equation}
The second limit on the right hand side of (\ref{ld}) vanishes, because
\[
0 \leq \int_{|\lambda| > 1/s}  \frac{\sin(\lambda s)}{s}\; d\mu(\lambda) \leq \frac{1}{s}\int_{|\lambda| > 1/s} d\mu(\lambda) \to 0 ,
\]
while the first limit equals  $+\infty$, because
\[
\sin 1 \int_{|\lambda|\leq 1/s} \lambda \; d\mu(\lambda) \leq \int_{|\lambda|\leq 1/s} \frac{\sin(\lambda s)}{s}\; d\mu(\lambda),
\]
and
\[
\lim_{s \to 0^+} \int_{|\lambda|\leq 1/s} \lambda \; d\mu(\lambda) = \int_{\RM} \lambda\; d\mu(\lambda)=+\infty.
\]
Thus
\[
\lim_{s \to 0^+} \Im \mathcal{A}'(s)=-\lim_{s \to 0^+}\int_{\RM}  \frac{\sin(\lambda s)}{s}\; d\mu(\lambda)  = -\infty.
\]
Similarly, one can prove that
\[
\lim_{s \to 0^-} \Im \mathcal{A}'(t)= + \infty.
\]
\qed
\end{example}

Note that in Theorem \ref{iff condition QZE multidimensional} there is no mention to a lower bound of the Hamiltonian. Lower boundedness is something of red herring.
It has played a crucial role in QZD; in fact, it has been always advocated in the literature, and indeed the limiting (Zeno) Hamiltonian which engenders the effective dynamics is nothing but the Friedrichs extension of $PHP$.
However, if one is concerned with the QZE \textit{per se},
such hypothesis is quite unnatural. Therefore, one can wonder whether lower-boundedness is really a physical requirement for QZD, or rather it is just a --very convenient-- technical hypothesis. Our second main result answers this question. It gives a characterization of the quantum Zeno dynamics in which lower boundedness plays no role.

\begin{theorem}\label{thm:QZE}
Consider  a self-adjoint operator  $H$ and an orthogonal
finite-rank projection $P$ in a complex Hilbert space
$\mathcal{H}$. Let  $\{P_{\Omega}^H\}$ be the projection-valued
spectral measure  of $H$ and $\{U(t)=e^{-itH}\}_{t \in \RM}$ the
one-parameter unitary group generated by $H$. Consider the product
formula
$V_{N}(t)=(PU(t/N)P)^{N}$, with $t \in \RM$ and $N \in \NM^*$ and
 the family of self-adjoint operators $H^{(\Lambda)}=HP^{H}_{(-\Lambda,\Lambda)}$, with $\Lambda > 0$.
The following statements are equivalent:
\begin{enumerate}
    \item \label {thm:QZE_1}
    \begin{equation*}
     P\, P_{(-\Lambda,\Lambda)^c}^H\, P = o\left( \frac{1}{\Lambda} \right), \qquad \mathrm{for} \quad \Lambda \to + \infty,
    \end{equation*}
    and there exists bounded the limit
    $$H_Z=\lim_{\Lambda \to +\infty} PH^{(\Lambda)}P;$$
     \item
      \begin{equation*}
       \left.\frac{d}{ds}V(s)\right|_{s=0}=-iH_{Z};
      \end{equation*}
    \item
    \begin{equation*}
      \lim_{N \to +\infty} V_{N}(t)=Pe^{-itH_{Z}},
    \end{equation*}
    uniformly for $t$ in finite intervals of $\RM$.
\end{enumerate}
\end{theorem}

Therefore,  the existence of QZD is equivalent  to
the energy fall-off property, which assures the existence of QZE, \emph{and}
to the existence of a limit mean value Hamiltonian $H_Z = \lim_\Lambda P H^{(\Lambda)}P$.

\begin{remark}
 In the one-dimensional case, when $P=|\psi\rangle\langle\psi|$, by using Eq.~(\ref{eq:VsAs}), the QZD product formula (\ref{ampl. form}) reads
\begin{equation*}
V_N(t)=[\mathcal{A}(t/N)]^N P.
\end{equation*}
In such a case QZD is trivial and its existence is equivalent to the existence of the numerical limit
\begin{equation}
\lim_{N\to+\infty} [\mathcal{A}(t/N)]^N = e^{-i t E_Z},
\label{eq:qze1dim}
\end{equation}
with a finite phase $E_Z \in \RM$. This has to be compared with  QZE, where one looks at the modulus of Eq.~(\ref{eq:qze1dim}), and thus the existence of a finite mean energy is not necessary, as shown also in Example~\ref{ex:part}.

Theorem~\ref {thm:QZE} states that the phase $E_Z$ is finite iff the limit
\begin{equation*}
H_Z = \lim_{\Lambda\to+\infty} (\psi,H^{(\Lambda)}\psi) P,
\end{equation*}
exists bounded, and in such a case one has
\begin{equation*}
E_Z = \lim_{\Lambda\to+\infty} (\psi,H^{(\Lambda)}\psi).
\end{equation*}
\qed
\end{remark}

From Theorem~\ref {thm:QZE} one immediately gets that, if the attention is restricted to  positive Hamiltonians, the condition given in Theorem~\ref{th:EI}  on the domain of the square root is both necessary and sufficient. Indeed, we have
\begin{corollary}\label{cor:QZE}
Let  $H$ be a
positive self-adjoint operator 
and $P$ be an orthogonal finite-rank projection onto $\mathcal{H}_{P}=P\mathcal{H}$.
Then,
\[
\mathcal{H}_{P} \subset D(H^{1/2})\quad \Leftrightarrow \quad  \lim_{N \to +\infty} V_{N}(t)= P\exp\left(-it(H^{1/2}P)^{*}(H^{1/2}P)\right) .
\]
\end{corollary}
\begin{proof}

One implication is the content of Theorem \ref{th:EI}. The other follows by Theorem \ref{thm:QZE} after noting that, when $H\geq 0$,
\begin{eqnarray*}
P H^{(\Lambda)} P= P \int_{[0,\Lambda)} \lambda\; dP_\lambda^H P,
\end{eqnarray*}
and thus the existence of a bounded limit  $\lim_{\Lambda} P H^{(\Lambda)} P$ implies that $\| H^{1/2}P\|<\infty$.
\end{proof}

\begin{remark}
By looking at Corollary \ref{cor:QZE}, one might think that the results for positive operators hold true in the general unbounded case by replacing the condition $\mathcal{H}_{P} \subset D(H^{1/2})$ with $\mathcal{H}_{P} \subset D(|H|^{1/2})$. Unfortunately, this is not true. The condition $\mathcal{H}_{P} \subset D(|H|^{1/2})$ is stronger than Condition~\ref{thm:QZE_1}.\ in Theorem~\ref{thm:QZE}, and in fact is a sufficient condition for QZD, but is not necessary. Indeed, it is easy to construct a probability Borel measure $\mu_{\psi}$ associated to the Hamiltonian $H$ at some $\psi \in \mathcal{H}$ such that
$$
\lim_{\Lambda \to +\infty} \int_{(-\Lambda,\Lambda)} \lambda \; d\mu_{\psi}(\lambda) <+\infty,
$$
while
$$
\lim_{\Lambda \to +\infty} \int_{(-\Lambda,\Lambda)} |\lambda| \; d\mu_{\psi}(\lambda) =+\infty.
$$
Observe that in this case if one considers the projection $P=|\psi\rangle \langle \psi|$, one gets that $\mathcal{H}_{P} \not\subset D(|H|^{1/2}) $, despite the fact that the pair  projection-Hamiltonian $(P,H)$ satisfies Statement~\ref{thm:QZE_1}.\ of Theorem~\ref{thm:QZE}. \qed
\end{remark}

\section{\label{sec:iff condition}Proofs of the theorems}

Let us now turn to the proofs of our characterizations of the quantum Zeno
effect and its dynamics, Theorems \ref{iff condition QZE multidimensional} and \ref{thm:QZE}.
First of all let us prove a preliminary lemma that will be useful in the following. We note, incidentally, that this Tauberian result is interesting in itself and is probably known in the probability community. However, we will give here a purely analytical proof.
\begin{lemma}\label{tauberian lemma}
Let $\mu$ be a Borel measure on $\RM$ with $\mu(\RM)=1$. 
The following assertions are equivalent:
\begin{enumerate}
    \item
    \begin{equation*}
      \mu((-\Lambda,\Lambda)^c) = o\left( \frac{1}{\Lambda} \right), \quad \mathrm{for} \quad \Lambda \to + \infty
    \end{equation*}
    \item
      \begin{equation*}
        \frac{1}{\Lambda^{k+1}}\int_{(-\Lambda,\Lambda)} \lambda^{k+1}\; d\mu(\lambda) = o\left( \frac{1}{\Lambda} \right), \quad \mathrm{for} \quad \Lambda \to + \infty, \quad
        \textrm{for every $k \in \NM^*$}.
      \end{equation*}
    \end{enumerate}
\end{lemma}

\begin{proof}
\quad\newline

$1. \Rightarrow 2.$

Let $k \in \NM^*$, then for every $\Lambda >0$, by using an integration by parts formula, see e.g.\ \cite{Riesz}, we have that
\begin{eqnarray*}
& &\frac{1}{\Lambda^{k}} \int_{(-\Lambda,\Lambda)} \lambda^{k+1} d\mu(\lambda) \nonumber\\
& & =  \frac{1}{\Lambda^{k}} \Big[ \Lambda^{k+1} \mu\left((-\infty,\Lambda]\right)-(-\Lambda)^{k+1}\mu\left((-\infty,-\Lambda]\right)
\nonumber\\
& &\quad\qquad -(k+1) \int_{(-\Lambda,\Lambda)}\lambda^{k} \mu ((-\infty, \lambda]) \; d \lambda \Big] \nonumber\\
& & = \Lambda \mu\left((-\infty,\Lambda]\right)+(-1)^{k}\Lambda\mu\left((-\infty,-\Lambda]\right) \nonumber \\
& & \quad -\frac{k+1}{\Lambda^{k}} \int_{(0,\Lambda)}\lambda^{k} \left(\mu ((-\infty, \lambda])+(-1)^{k}\mu((-\infty, -\lambda])\right) \; d \lambda.
\end{eqnarray*}
We can write
\begin{eqnarray}
& &\frac{1}{\Lambda^{k}} \int_{(-\Lambda,\Lambda)} \lambda^{k+1} d\mu(\lambda) \nonumber\\
& & = \Lambda \mu \left( (-\Lambda,\Lambda] \right)-\frac{k+1}{\Lambda^k}\int_{(0,\Lambda)}\lambda^k \mu\left((-\lambda,\lambda]\right) \; d \lambda \nonumber\\
&  &\quad + \left(1+(-1)^k\right)\left(\Lambda \mu \left( (-\infty,-\Lambda] \right)-\frac{k+1}{\Lambda^k}\int_{(0,\Lambda)}\lambda^k \mu\left((-\infty,-\lambda] \right) \; d \lambda \right).
\label{int. part}
\end{eqnarray}
The second line of (\ref {int. part}) reads
\begin{eqnarray*}
&  & \Lambda \Big[1-\mu \left( (-\Lambda,\Lambda]^c \right) \Big] -\frac{k+1}{\Lambda^k}\int_{(0,\Lambda)}\lambda^k \Big( 1-\mu ((-\lambda,\lambda]^c)\Big) \; d \lambda \nonumber\\
& = & 
- \Lambda \mu \left( (-\Lambda,\Lambda]^c \right)
+ \frac{k+1}{\Lambda^k} \int_{(0,\Lambda)}\lambda^k \mu ((-\lambda,\lambda]^c) \; d \lambda \nonumber\\
& \leq & - \Lambda \mu ( (-\Lambda,\Lambda]^c) + \frac{k+1}{\Lambda} \int_{(0,\Lambda)}\lambda \mu ((-\lambda,\lambda]^c) \; d \lambda  \to 0, \quad \textrm{for} \quad \Lambda \to +\infty,
\end{eqnarray*}
while in the third line of (\ref {int. part}), which is nonzero only for $k$ even, one gets
$$
\Lambda \mu \left( (-\infty,-\Lambda] \right) \leq \Lambda \mu \left( (-\Lambda,\Lambda)^c \right) \to 0,
$$
and
$$
\frac{1}{\Lambda^k}\int_{[0,\Lambda]}\lambda^k \mu\left((-\infty,-\lambda]\right) \; d \lambda
\leq \frac{1}{\Lambda}\int_{[0,\Lambda]}\lambda \mu\left((-\lambda,\lambda)^c\right) \; d \lambda
\to 0.
$$
Therefore, we have that
$$
\lim_{\Lambda \to +\infty} \frac{1}{\Lambda^{k}} \int_{(-\Lambda,\Lambda)} \lambda^{k+1} d\mu(\lambda)=0.
$$

$2. \Rightarrow 1.$

Let us choose $k=1$ and fix $\epsilon >0$. By hypothesis we have that there exists a real $\Lambda_{0}>0$ such that for every $\Lambda>\Lambda_{0}$
$$
\frac{1}{\Lambda}\int_{(-\Lambda,\Lambda)} \lambda^2 \; d\mu(\lambda) <\epsilon.
$$
Thus, for every $\Lambda>\Lambda_{0}$,
\begin{eqnarray*}
\Lambda\int_{(-\Lambda,\Lambda)^c} d\mu(\lambda) & = & \Lambda \sum_{k=0}^{+\infty} \int_{2^{k}\Lambda \leq |\lambda| < 2^{k+1}\Lambda} d\mu(\lambda) \\
& \leq & \Lambda \sum_{k=0}^{+\infty}   \int_{2^{k}\Lambda \leq |\lambda| < 2^{k+1}\Lambda} \left(\frac{\lambda}{2^{k}\Lambda}\right)^2 d\mu(\lambda) \\
                                                    & \leq & \sum_{k=1}^{+\infty} \frac{1}{2^{k-1}} \frac{1}{2^{k+1}\Lambda} \int_{|\lambda| < 2^{k+1}\Lambda } \lambda^2 d\mu(\lambda) \\
                                               & \leq &  2\epsilon .
\end{eqnarray*}
\end{proof}

\begin{remark}
\label{only k}
Observe that in order to prove that $2. \Rightarrow 1.$ it is sufficient that
$$
\frac{1}{\Lambda^3}\int_{(-\Lambda,\Lambda)} \lambda^{2}\; d\mu(\lambda) = o\left( \frac{1}{\Lambda} \right), \quad \textrm{for $\Lambda \to + \infty$}.
$$
\qed
\end{remark}

Now we prove  Proposition \ref{iff condition QZE} on the characterization of the one-dimensional QZE. We will use it as the first step in the proof of the multi-dimensional case, Theorem \ref{iff condition QZE multidimensional}.

\subsection*{Proof of Proposition~\ref {iff condition QZE}}

Let us start with the proof of the first equivalence $1. \Leftrightarrow 2.$\\
Observe that $\mathcal{A}$ is a (uniformly) continuous function and that $\mathcal{A}(0)=1$. Define for every $s \in \RM$
$$
z(s)=\mathcal{A}(s)-1\,,
$$
so that $z$ is a continuous function with $z(0)=0$.
Recall that
\begin{equation}\label{rel dA e dp}
p'(0)=\lim_{s \to 0} 2 \; \frac{\Re (z(s))}{s}
\end{equation}
and
$$
\frac{\Re (z(s))}{s} =\frac{1}{s} \int_{\RM} \left(\cos(\lambda s)-1\right)\; d\mu(\lambda)=-\frac{2}{s} \int_{\RM} \sin^2 \left(\frac{\lambda s}{2}\right) \; d\mu(\lambda) .
$$
We can write
\begin{eqnarray*}
\frac{2}{|s|} \int_{\RM} \sin^2 \left(\frac{\lambda s}{2}\right)\; d\mu(\lambda)
= g(s)+h(s) ,
\end{eqnarray*}
where
$$
g(s)=\frac{2}{|s|} \int_{|\lambda|<2/|s|} \sin^2 \left(\frac{\lambda s}{2}\right) \; d\mu(\lambda)
$$
and
$$
h(s)=\frac{2}{|s|} \int_{|\lambda|\geq 2/|s|} \sin^2 \left(\frac{\lambda s}{2}\right)\; d\mu(\lambda) .
$$
Therefore, since $g,h \geq 0$, one has that
\begin{equation}\label{equiv. deriv. 0}
\lim_{s \to 0} 2 \;\frac{\Re (z(s))}{s}=0 \quad \Leftrightarrow \quad \lim_{s \to 0} g(s)=\lim_{s \to 0} h(s)=0.
\end{equation}
Observe that, since $x^2 \sin^2 1 \leq \sin^2 x \leq x^2$ for $|x|<1$,
$$
\sin^2 1 \frac{|s|}{2} \int_{|\lambda|< 2/|s|} \lambda^2 \;d\mu(\lambda) \leq g(s) \leq \frac{|s|}{2} \int_{|\lambda|< 2/|s|}\lambda^2 \;d\mu(\lambda) .
$$
Therefore
\begin{equation}\label{sin2 as l2}
\lim_{s \to 0} g(s) = 0 \quad \Leftrightarrow \quad  \lim_{s \to 0} \frac{|s|}{2} \int_{|\lambda|< 2/|s|} \lambda^2 \;d\mu(\lambda)=0 .
\end{equation}

$1. \Rightarrow 2.$

Using Lemma~\ref{tauberian lemma} and (\ref{sin2 as l2}) one gets that $g(s) \to 0$, for $s \to 0$. Moreover,
$$
0 \leq h(s) \leq \frac{2}{|s|}\mu\left((-2/|s|,2/|s|)^c\right)\to 0, \quad s\to 0.
$$

$2. \Rightarrow 1.$

Observe that, using (\ref{equiv. deriv. 0}), we have that
$$
p'(0)=0 \Rightarrow \lim_{s \to 0} g(s)=0.
$$
Thus, by using (\ref{sin2 as l2}) and Remark \ref{only k} we prove the thesis.

Now we prove the second equivalence $2. \Leftrightarrow 3.$\\
Observe that
\begin{equation}
\label{eq:pvsS}
[p(t/N)]^N -1  =  \sum_{k=0}^{N-1}[p(t/N)]^k \left(p(t/N)-1\right)= S_{N}(t)N\left(p(t/N)-1\right),
\end{equation}
where
$$
S_{N}(t)= \frac{1}{N}\sum_{k=0}^{N-1}[p(t/N)]^k.
$$

$2. \Rightarrow 3.$

Since $0 \leq S_{N}(t) \leq 1$, we have from (\ref{eq:pvsS}) that
$$
\left|[p(t/N)]^N -1 \right| \leq N \left|p(t/N)-1\right| \to 0, \quad N \to +\infty,
$$
uniformly for $t$ in finite intervals of $\RM$.

$3. \Rightarrow 2.$

We know that $[p(t/N)]^N \to 1$ for $N \to +\infty$ uniformly in $t$ in finite intervals of $\RM$.
Observe that, since $0\leq p(t/N)\leq 1$,
$$
1 \geq S_{N}(t) \geq [p(t/N)]^N \to 1,
$$
uniformly for $t$ in finite intervals of $\RM$. Therefore, from (\ref{eq:pvsS}),
$$
\lim_{N \to +\infty} N\left(p(t/N)-1\right)=0,
$$
and thus $p'(0)=0$.
\qed

Now that we have gathered all necessary ingredients, let us conclude this section with the proofs of our main results,  Theorems \ref{iff condition QZE multidimensional} and \ref{thm:QZE}.

\subsection*{Proof of Theorem~\ref{iff condition QZE multidimensional}}

$1. \Rightarrow 2.$ 

From $Z_1(s)=V(s)^{*}V(s)$ one gets
\[
Z'_1(s)=\left(\frac{d}{ds}V(s)^{*}\right)V(s)+V(s)^{*}\left(\frac{d}{ds}V(s)\right).
\]
Therefore, for every $\phi \in \mathcal{H}$, we have
\begin{eqnarray*}
(\phi,Z'_1(0)\phi) =  \frac{d}{ds} \left[\left( e^{isH}P\phi, P\phi\right) +\left( e^{-isH}P\phi, P\phi \right)\right]_{s=0} .
\end{eqnarray*}
If $P\phi\neq 0$, let us define the probability Borel measure on $\RM$
\[
d\mu(\lambda)=\frac{1}{\|P \phi\|^2}d(P\phi,P_{\lambda}^H P\phi)
\]
and the survival amplitude
\[
\mathcal{A}(s)=\int_{\RM}e^{-is\lambda} \; d\mu (\lambda).
\]
Therefore,
\begin{eqnarray}\label{addends}
(\phi,Z'_1(0)\phi) =2 \|P \phi\|^2 \left.\frac{d}{ds}\Re \left( \mathcal{A}(s) \right)  \right|_{s=0} .
\end{eqnarray}
By Condition \ref{fdtc}.\ we get that $\mu$ satisfies
Condition~\ref{fdtc 1D}.\ of Proposition~\ref{iff condition QZE}. Therefore, the right side of
(\ref{addends}) vanishes and, by the polarization identity, it follows that $Z'_1(0)=0$.

$2. \Rightarrow 1.$ 

Let $\psi \in \mathcal{H}$, $\| P \psi \|= 1$, and consider the
Borel probability measure
\[
d \mu (\lambda)= d(P \psi , P_{\lambda}^H P \psi).
\]
Define for every $s \in \RM$
\[
\mathcal{A}(s)=\int_{\RM} e^{-is \lambda}\; d\mu(\lambda) \quad
\textrm{and} \quad p(s)=|\mathcal{A}(s)|^2.
\]
Observe that
\[
p'(0)=(P\psi, Z'_1(0) P \psi)=0,
\]
thus, using the equivalence proved in Proposition \ref{iff condition
QZE}, we have
\[
\lim_{\Lambda \to + \infty} \Lambda \int_{(-\Lambda,\Lambda)^c}\; d\mu(\lambda)=\lim_{\Lambda \to + \infty} \Lambda (\psi,PP^{H}_{(-\Lambda,\Lambda)^c}P\psi)=0.
\]
Since $\mathcal{H}_{P}$ is a finite dimensional space we have proved Condition \ref{fdtc}.

$2. \Rightarrow 3.$

Use  the telescopic sum:
\begin{eqnarray}
Z_N(t)-P &=& V_{N}(t)^{*}V_{N}(t)-P
\nonumber \\
                          & = & \sum_{k=0}^{N-1}\Big[V\Big(\frac{t}{N}\Big)^*\Big]^k \Big[V\Big(\frac{t}{N}\Big)^*V\Big(\frac{t}{N}\Big)-P\Big]\Big[V\Big(\frac{t}{N}\Big)\Big]^{N-1-k} \nonumber\\
                          & = & \sum_{k=0}^{N-1}\Big[V\Big(\frac{t}{N}\Big)^*\Big]^k \Big[Z_1\Big(\frac{t}{N}\Big)-P \Big]\Big[V\Big(\frac{t}{N}\Big)\Big]^{N-1-k}.
\label{telescopic}
\end{eqnarray}
Therefore, since $\| V(t/N)\|\leq 1$, one gets
\[
\left\|Z_{N}(t)-P\right\| \leq N \left\|
Z_1(t/N) -P \right\| \to  0 ,
\]
uniformly for $t$ in finite intervals of $\RM$ by hypothesis.

$3. \Rightarrow 2.$

We want to prove that
\begin{equation}\label{deriv2}
\lim_{N \to + \infty} N(Z_1(t/N)-P)=
\lim_{N \to + \infty} N(V(t/N)^* V(t/N)-P)=0,
\end{equation}
uniformly for $t$ in finite intervals of $\RM$. Observe that
(\ref{telescopic}) can be expanded also in this way
\begin{eqnarray}
Z_N(t)-P
&=&  N \left[
V\Big(\frac{t}{N}\Big)^* S_N(t) V\Big(\frac{t}{N}\Big) - S_N(t) \right],
\label{telescopic2}
\end{eqnarray}
where
\begin{equation*}
S_N(t)=\frac{P}{N}\sum_{k=0}^{N-1}\Big[V\Big(\frac{t}{N}\Big)^*\Big]^k \Big[V\Big(\frac{t}{N}\Big)\Big]^k.
\end{equation*}
Let us first prove that the  ergodic sum $S_N(t)$ tends to $P$,
\[
\lim_{N \to +\infty} S_N(t)=P,
\]
uniformly for $t$ in finite intervals of $\RM$.  It is easy to see that
for every $k\geq l\geq 0$
\[
\Big[V\Big(\frac{t}{N}\Big)^*\Big]^k \Big[V\Big(\frac{t}{N}\Big)\Big]^k \leq \Big[V\Big(\frac{t}{N}\Big)^*\Big]^l \Big[V\Big(\frac{t}{N}\Big)\Big]^l ,
\]
whence, for every $k \in \{ 0,1, \ldots, N-1\}$,
\[
0 \leq P- P \Big[V\Big(\frac{t}{N}\Big)^*\Big]^k \Big[V\Big(\frac{t}{N}\Big)\Big]^k \leq P- Z_N(t) .
\]
Therefore,
\begin{eqnarray*}
0 \leq P- S_N(t) =\frac{1}{N} \sum_{k=0}^{N-1}\Big( P- P \Big[V\Big(\frac{t}{N}\Big)^*\Big]^k \Big[V\Big(\frac{t}{N}\Big)\Big]^k\Big) \leq P- Z_N(t) \to 0,
\end{eqnarray*}
by hypothesis.

Assume that (\ref{deriv2}) is not valid. By taking the trace of (\ref{telescopic2}) and by using its cyclic property we get
\begin{equation*}
\tr(Z_N(t)-P )
=   \tr \left[ N\Big(
V\Big(\frac{t}{N}\Big) V\Big(\frac{t}{N}\Big)^* - P \Big) S_N(t) \right].
\end{equation*}
Since the ergodic sum $S_N(t)$ is a positive operator whose limit is $P$, the right hand side does not tend to 0, while the left hand side vanishes by hypotesis, and we get a contradiction.
\qed

\subsection*{Proof of Theorem~\ref{thm:QZE}}

Let us start with the proof of the first equivalence $1. \Leftrightarrow 2.$\\
Let $\Re V(s)= (V(s)+V(s)^*)/2$ and $\Im V(s)= (V(s)-V(s)^*)/2 i$ for all $s \in \RM$.
Observe that by Theorem~\ref{iff condition QZE multidimensional} it follows that
\begin{equation}\label{equiv.measure.tail}
P\,P_{(-\Lambda,\Lambda)^c}^H\, P = o\left( \frac{1}{\Lambda} \right) \Leftrightarrow \left.\frac{d}{ds}V(s)^* V(s)\right|_{s=0}=2\left.\frac{d}{ds}\Re  V(s)\right|_{s=0}=0.
\end{equation}
Now we prove that
\begin{equation}\label{equiv imm}
H_{Z}=\lim_{\Lambda \to +\infty} PH^{(\Lambda)}P
\Leftrightarrow -\left.\frac{d}{ds}\Im V(s)\right|_{s=0}=H_{Z}.
\end{equation}
Let us denote $dQ_{\lambda}=PdP^{H}_{\lambda}P$.
Observe that
\begin{eqnarray*}
-\left.\frac{d}{ds} \Im V(s)\right|_{s=0} & = & \lim_{s \to 0} \frac{1}{s}P\sin(sH)P
                            = \lim_{s \to 0} \int_{\RM} \frac{\sin(\lambda s)}{s}\;dQ_{\lambda}
\end{eqnarray*}
In order to prove (\ref{equiv imm}) we will prove that
\begin{equation}\label{lim diff}
\lim_{s \to 0} PH^{(\pi/s)}P-\frac{1}{s}P\sin(sH)P= 0.
\end{equation}
We have that, if $s>0$
\begin{eqnarray*}
& & PH^{(\pi/s)}P-\frac{1}{s}P\sin(sH)P \\
& & = \int_{(-\pi/s,\pi/s)} \lambda \;dQ_{\lambda} - \frac{1}{s} \int_{\RM} \sin(\lambda s)\; dQ_{\lambda} \\
& & = \int_{(-\pi/s,\pi/s)} \lambda \left(1-\frac{\sin(\lambda s)}{\lambda s}\right)\; dQ_{\lambda}-\frac{1}{s} \int_{(-\pi/s, \pi/s)^c} \sin(\lambda s)\; dQ_{\lambda} .
\end{eqnarray*}
Therefore, since $1-\sin x/ x\geq 0$, we get
\begin{eqnarray*}
& & -\frac{\pi}{s} \int_{(-\pi/s,\pi/s)} \left(1-\frac{\sin(\lambda s)}{\lambda s}\right)\; dQ_{\lambda}-\frac{1}{s} \int_{(-\pi/s, \pi/s)^c}  dQ_{\lambda} \\
& & \leq  PH^{(\pi/s)}P-\frac{1}{s}P\sin(sH)P \\
& & \leq \frac{\pi}{s} \int_{(-\pi/s,\pi/s)} \left(1-\frac{\sin(\lambda s)}{\lambda s}\right)\; dQ_{\lambda}+\frac{1}{s} \int_{(-\pi/s, \pi/s)^c} dQ_{\lambda}.
\end{eqnarray*}
By noting that
$$
0 \leq \frac{\pi}{s}\int_{(-\pi/s, \pi/s)}\left(1-\frac{\sin(\lambda s)}{\lambda s}\right)\; dQ_{\lambda} \leq \frac{\pi s}{6} \int_{(-\pi/s, \pi/s)} \lambda^2 \; dQ_{\lambda}
$$
and by using (\ref{equiv.measure.tail}) and Lemma \ref{tauberian lemma} we obtain that (\ref{lim diff}) holds when $s\to 0^+$.
With the same argument one can prove the thesis when $s\to 0^-$.

$2. \Rightarrow 3.$

Observe that
\begin{eqnarray*}
\left\|V_{N}(t)-Pe^{-itH_{Z}}\right\| & = & \left\|\left(V(t/N)\right)^N-(Pe^{-itH_{Z}/N})^N\right\| \\
                                      & = & \left\| \sum_{k=0}^{N-1} (V(t/N))^{N-1-k} (V(t/N)-Pe^{-itH_{Z}/N})Pe^{-iktH_{Z}/N} \right\| \\
                                      & \leq & N \left\|V(t/N)-Pe^{-itH_{Z}/N}\right\| \to 0
\end{eqnarray*}
uniformly for $t$ in finite intervals of $\RM$.

$3. \Rightarrow 2.$

Let $z>0$. We will prove that
$$
\lim_{N \to +\infty}\left( z-N \left(V(t/N)-P\right)\right)^{-1}P =(z+itH_{Z})^{-1}P,
$$
uniformly for $t$ in finite intervals of $\RM$. This implies that
$$
\lim_{N \to +\infty} N(V(t/N)-P) = -iH_{Z}
$$
uniformly for $t$ in finite intervals of $\RM$, and thus
$$
\left.\frac{d}{ds}V(s)\right|_{s=0}=-iH_{Z}.
$$
Indeed, observe that
\begin{eqnarray}\label{3.2 res}
\left( z-N\left(V(t/N)-P\right)\right)^{-1}P  =  \frac{1}{N}\sum_{k=0}^{+\infty}\frac{V(t/N)^k}{(1+z/N)^{k+1}}P  =  \int_{0}^{+\infty}\frac{V(t/N)^{[sN]}}{(1+z/N)^{[sN]+1}} P \; ds ,
\end{eqnarray}
where $[\cdot]$ denotes the integer part function.
By the dominated convergence theorem, the right hand side of (\ref{3.2 res}) converges to
$$
\int_{0}^{+\infty} e^{-sz} Pe^{-istH_{Z}}P \;  ds = (z+itH_{Z})^{-1}P,
$$
uniformly for $t$ in finite intervals of $\RM$.
\qed

%\subsection{Subsection title}
%\label{sec:2}
%as required. Don't forget to give each section
%and subsection a unique label (see Sect.~\ref{sec:1}).
%%
%% For one-column wide figures use
%\begin{figure}
%% Use the relevant command for your figure-insertion program
%% to insert the figure file.
%% For example, with the option graphics use
%\centering
%\resizebox{0.75\textwidth}{!}{%
%  \includegraphics{leer.eps}
%}
%% If not, use
%%\vspace{5cm}       % Give the correct figure height in cm
%\end{figure}
%\begin{table}
%\centering
%\begin{tabular}{lll}
%\hline\noalign{\smallskip}
%\end{tabular}
%% Or use
%%\vspace*{5cm}  % with the correct table height
%\end{table}
%
\acknowledgments
We would  like to thank Sandro Graffi,
Andrzej Kossakowski, Hiromichi Nakazato, Saverio Pascazio and Shuichi Tasaki  for stimulating
discussions.
%
% BibTeX users please use
% \bibliographystyle{}
% \bibliography{}

\end{document}